\documentclass[prl,twocolumn,floatfix,superscriptaddress]{revtex4-1}
\usepackage{dcolumn,amsmath}
\usepackage{graphicx}
\usepackage{bm}
\usepackage{hyperref}
\usepackage{xcolor}

\usepackage[utf8]{inputenc}
\usepackage[T1]{fontenc}

\usepackage{tabularx}
\usepackage{braket}

\begin{document}

\title{Theoretical molecular spectroscopy of actinide compounds: The ThO molecule}

\author{Andr\'ei V. Zaitsevskii}
\affiliation{Petersburg Nuclear Physics Institute named by B.P.\ Konstantinov of National Research Center ``Kurchatov Institute'' (NRC ``Kurchatov Institute'' - PNPI), 1 Orlova roscha, Gatchina, 188300 Leningrad region, Russia}
\affiliation{Department of Chemistry, M.V. Lomonosov Moscow State University, Leninskie gory 1/3, Moscow, 119991~Russia}

\author{Alexander V. Oleynichenko}
\email{oleynichenko\_av@pnpi.nrcki.ru \\ alexvoleynichenko@gmail.com}
\affiliation{Petersburg Nuclear Physics Institute named by B.P.\ Konstantinov of National Research Center ``Kurchatov Institute'' (NRC ``Kurchatov Institute'' - PNPI), 1 Orlova roscha, Gatchina, 188300 Leningrad region, Russia}
\homepage{http://www.qchem.pnpi.spb.ru}

\author{Ephraim Eliav}
\affiliation{School of Chemistry, Tel Aviv University, 6997801 Tel Aviv, Israel}

\begin{abstract}
The tiny-core generalized (Gatchina) relativistic pseudopotential (GRPP) model provides an accurate approximation for many-electron Hamiltonians of molecules containing heavy atoms, ensuring a proper description of the effects of non-Coulombian electron-electron interactions, electronic self-energy and vacuum polarization. Combining this model with electron correlation treatment in the frames of the intermediate Hamiltonian Fock space coupled cluster theory employing incomplete main model spaces, one obtains a reliable and economical tool for excited state modeling. The performance of this method is assessed in applications to \emph{ab initio} modeling of excited electronic states of the thorium monoxide molecule with term energies below 20 000 cm$^{-1}$. Radiative lifetimes of excited states are estimated using truncated expansions of effective and metric operators in powers of cluster amplitudes.
\end{abstract}

\maketitle

\section{Introduction}\label{sec:intro}

Up to now, molecular systems containing actinide atoms remain a challenge for \textit{ab initio} modeling (\cite{Pepper:91,Gagliardi:07,Dolg:15,Kovacs:15,Kovacs:20} and references therein). There are several reasons for such a disappointing situation. First of all, in actinide compounds, both relativistic and correlation effects are powerful and intertwined with each other. It is well-established that 
a
proper treatment of relativistic effects 
involves the incorporation of
two-electron Breit interaction~\cite{Petrov:04}. Moreover, the accounting for QED effects is important for high-precision modeling of spectroscopic properties
\cite{Oleynichenko:LIBGRPP:23}. Furthermore, most actinide compounds possess several open shells 
resulting in highly dense spectra of electronic states and strong static correlations, which can be handled within single-reference electronic structure methods only in exceptional cases. The problems mentioned make relativistic multi-reference models like configuration interaction (MR-CI)~\cite{Fleig:06}, coupled cluster (MR-CC)~\cite{Visscher:01,Ghosh:16,Oleynichenko:CCSDT:20,Eliav:Review:22} or 
quasidegenerate
many-body perturbation theory based methods~\cite{Dzuba:96,Zaitsevskii:RMPPT:02,Abe:06,Savukov:21} the preferable choice to deal with such systems.

At the same time, theoretical supply is highly demanded by experimenters, first of all working in the field of high-resolution spectroscopy of short-lived radioactive molecules, rapidly growing in the last few years~\cite{GarciaRuiz:20,ArrowsmithKron:23,Udrescu:23}. First-principle modeling allows one 
to plan new experiments~\cite{AcF:Proposal:21,Zaitsevskii:RaF:22}
as well as to interpret the obtained data. For instance, only high-precision electronic structure calculation can provide electronic factors used to set limits for the $\mathcal{P}$,$\mathcal{T}$-violating interactions~\cite{Safronova:18,Alarcon:22}. The other fields which are inconceivable without intensive theoretical support include direct laser cooling~\cite{Kozyryev:17,Isaev:16,Isaev:Review:18,Ivanov:19} and laser assembly of cold molecules~\cite{Pazyuk:15,Fleig:21,Klos:22}. In this regard, the most important molecular properties which the theory can provide are excitation energies, spectroscopic constants and radiative lifetimes of excited states. Furthermore, state-of-the-art experiments usually require 
the knowledge of
properties characterizing response to external electromagnetic fields, such as permanent molecule frame dipole moment.

The challenges for the theory described above make it very difficult to obtain accurate results useful for experimenters. However, intensive work is underway to overcome these difficulties. 
Recently~\cite{Zaitsevskii:QED:22}, 
a new formulation of the relativistic multireference coupled cluster theory for the Fock space (FS RCC) was proposed. This version of the FS RCC method operates with the concept of intermediate Hamiltonian (IH) reformulated for incomplete main model spaces to obtain smooth potential energy surfaces in some range of molecular geometry parameters. Such an approach solves the problem of very dense spectra and severe static correlation, at least for systems with up to two unpaired electrons. 
To bypass the necessity of four-component calculations with the Dirac-Coulomb-Breit Hamiltonian, which are highly demanding in the molecular framework, the generalized relativistic pseudopotential (GRPP) approach~\cite{Tupitsyn:95,Titov:99,Petrov:04,Mosyagin:16,Mosyagin:20,Zaitsevskii:QED:22,Wang:ccECP:22} has been successfully revived, and its accuracy was tested for 
actinide-containing systems (ThO and UO$_2$ molecules)~\cite{Oleynichenko:LIBGRPP:23}. GRPPs can also absorb both QED and finite-nuclear-size effects.

The evaluation of matrix elements of property operators is a long-standing problem in the coupled cluster theory~\cite{Monkhorst:77,Helgaker:12}. The main obstacles are the lack of an explicit expression for wavefunction and the non-variational nature of the theory, which restricts the use of the Hellman-Feynman theorem. The exponential Ansatz
used in FS RCC or other formulations of multireference CC, implies an infinite summation and thus gives only a recipe for calculating a wavefunction but not a wavefunction itself. For the special case of expectation value calculations using FS RCC, this problem 
is readily circumvented within the finite-difference approach~\cite{Abe:18,Haldar:21}. The other conceptually clear but technically complicated analytic approach is constructing the CC energy functional and solving state-specific equations for unknown Lagrange multipliers. Originally developed for the single-reference CC method~\cite{Salter:89,Gauss:91}, it was generalized to the case of multireference CC models~\cite{Szalay:95} and even implemented for non-relativistic FS CC in the $0h1p$, $1h0p$ and $1h1p$ sectors~\cite{Shamasundar:04,Ravichandran:11,Bhattacharya:14}. This analytic approach seems to be perfect for thoroughly study of molecular properties 
for a given electronic state. However, it becomes unreasonably expensive when several dozens of electronic states must be studied simultaneously, as it is normally required in spectra simulations for actinide compounds. A more 
practical method to calculate transition matrix elements 
should be oriented at obtaining data for all electronic states in a single calculation. One of such 
methods is the finite-field (FF) 
technique based on the approximate Hellmann-Feynman-like relation formulated for effective model-space operators, which leads to a very simple finite-difference formula for transition matrix elements~\cite{Zaitsevskii:Optics:18,Zaitsevskii:TDM:2020}. This method was shown to be pretty accurate 
for transition dipole moments~\cite{Krumins:20,Kruzins:21} and off-diagonal matrix elements of magnetic hyperfine interaction~\cite{Oleynichenko:HFS:20}. The main drawback of the FF technique in applications to highly symmetric molecules 
arises from the symmetry lowering by the perturbation operator. For example, calculations of transversal transition dipole moments in a diatomic heteronuclear molecule require passing from $C_{\infty v}$ to $C_s$ point group. An even more severe situation exists for the tensor parity nonconserving electron-nuclear interaction and for the operator representing the interaction between nuclear anapole moment and electronic subsystem, 
lowering the symmetry to the $C_1$ point group~\cite{Penyazkov:22}.

The latter consideration inspires the searches for alternative schemes aimed at obtaining all matrix elements simultaneously but maintaining high molecular symmetry. This is indeed possible within the framework of the theory of effective operators~\cite{Hurtubise:93}. The main idea of this approach consists of the direct use of the CC exponential wave operator in both bra- and ket-vectors with the subsequent truncation of the resulting infinite sum. 
The details depend on the specific definition of a property effective operator. Such a direct approach is well-established in atomic coupled cluster theory~\cite{Blundell:91,Safronova:99,Gopakumar:02}. It is widely used for high-precision calculations of transition dipole, magnetic dipole, and quadrupole matrix elements~\cite{Safronova:99,Sahoo:05,Safronova:17,Tan:23}, hyperfine interaction matrix elements~\cite{Porsev:06,Li:21}, parity non-conservation amplitudes~\cite{Sahoo:06,Porsev:10} and other quantities. To our best knowledge, this approach was not previously generalized to the case of molecular systems, except for its most straightforward and quite rough version in which cluster amplitudes are entirely neglected and the transition matrix element of property operator $O$ is approximated by its model space counterpart, $\braket{\psi_i|O|\psi_f} \approx \braket{\tilde{\psi}_i|O|\tilde{\psi}_f}$~\cite{Hehn:11,Zaitsevskii:TDM:2020}. In the present paper, we report the implementation of the direct scheme of transition moments calculations, including terms up to quadratic in cluster amplitudes for the Fock space sectors up to $0h2p$ (two electrons over the closed shell).

To assess the accuracy of all the novel techniques outlined above, it is appropriate to consider one of the simplest and, at the same time, quite typical actinide molecule, thorium monoxide (ThO). ThO is one of the most well-studied actinide-containing molecules since it is intensively used in experiments to detect the electron electric dipole moment ($e$EDM) conducted by the ACME collaboration~\cite{Vutha:10,Skripnikov:16,ACME:ThO:2018}. 
General features of low-lying electronic states of ThO were studied in the 1980s-2000s~\cite{Edvinsson:84,Edvinsson:85,Edvinsson:85jms,Goncharov:05}. Still, the most unique data on its molecule frame dipole moments and $g$-factors in different electronic states~\cite{Vutha:11,Wang:11,Kokkin:15,Wu:20}, as well as excited state lifetimes~\cite{Vutha:10,Kokkin:14,Wu:20,Ang:22} were obtained in the last decade in the framework of the preparation of $e$EDM experiments. Such a broad set of high-quality experimental data on molecular properties allows one to thoroughly assess the performance of relativistic electronic structure models aimed at applications in the field of theoretical spectroscopy. It is worth noting that excited states of ThO were previously studied by multireference perturbation theory MS-CASPT2~\cite{Paulovic:03} and relativistic Fock space coupled cluster method~\cite{Tecmer:18}. However, no calculations of properties, e.~g. radiative lifetimes, were presented. Furthermore, in~\cite{Tecmer:18} the Dirac-Coulomb (DC) Hamiltonian was employed, while it was recently shown in~\cite{Oleynichenko:LIBGRPP:23} that for ThO Gaunt interaction contributions to excitation energies reach 600 cm$^{-1}$, being comparable with vibrational frequencies. Thus the DC Hamiltonian cannot be regarded as reliable enough for this system if unambiguous vibrational numbering based on theoretical predictions is desired. 

The paper is organized in the following way. Firstly we recapitulate some features of the generalized relativistic pseudopotential model and recent development in the relativistic intermediate Hamiltonian Fock space coupled cluster method. Secondly, the new approach to calculate off-diagonal matrix elements between different electronic states in molecules is presented. Then the particular details of the computational procedure used in the present work are given, and calculated potential energy curves and spectroscopic constants, dipole moments, and excited state lifetimes of the ThO molecule are presented and compared with available experimental data. Finally, we draw conclusions about the scope of applicability of the presented models and discuss ongoing developments needed to further increase their accuracy and reliability.


\section{Theoretical considerations}\label{sec:theory}

\subsection{Tiny-core generalized relativistic pseudopotentials accounting for QED effects and Breit interaction}\label{sec:grpp}

To obtain the full picture of molecular properties interesting for spectroscopy one should be able to solve the electronic Schr\"{o}dinger equation (or its relativistic counterpart) for the set of electronic states $\psi_i$. In the most comprehensive molecular calculations to date many-electron wavefunctions $\psi_i$ are constructed from four-component one-particle functions and 
the Hamiltonian incorporates interelectronic zero-frequency Breit interactions and model Lamb shift operator 
\cite{Saue:11,Kelley:13,Sun:21,Sunaga:21,Sunaga:22,Eliav:Review:22,Hoyer:23}. Slightly more economical approximations are based on more or less accurate transformations to the two-component picture~\cite{Sikkema:09,Knecht:22}. These models are rather accurate, but their use would lead to prohibitively cumbersome computations even for moderate-size molecules. 

The practical solution is to pass to the relativistic pseudopotential (RPP) approximation~\cite{Titov:99,Schwerdtfeger:11,Dolg:12}. The basic idea of this approach is to use of some effective operator $H^{\rm RPP}$ instead of the ``exact'' relativistic Hamiltonian. In most cases, this operator also replaces some part of core electrons, thus greatly reducing the computational cost of the model. Moreover, it allows one to use the non-relativistic expression for kinetic energy and the ordinary Coulomb operator for two-electron interactions:
\begin{equation}
H^{\rm RPP} = \sum\limits_i
\left(
-\frac{\Delta_i}{2}
+\sum\limits_\alpha
\left(
-\frac{z_\alpha}{r_{\alpha i}}
+\hat{U}_{\alpha}(i)
\right)
\right)
+ \sum\limits_{i < k} \frac{1}{r_{ik}},
\end{equation}
where indices $i,k$ and $\alpha$ enumerate electrons and nuclei, respectively, $z_\alpha$ stands for the effective core charge (nuclear charge minus the number of electrons replaced by RPP), and $\hat{U}_\alpha$ denotes the RPP operator centered at nucleus $\alpha$. The latest generation of RPPs bears all information not only on the effects of relativity (including Breit~\cite{Stoll:02,Petrov:04}), but also effectively introduces finite nuclear size contributions~\cite{Mosyagin:20} and QED corrections (electron self-energy and vacuum polarization)~\cite{Hangele:12,Shabaev:13,Shabaev:18,Zaitsevskii:QED:22}.

To achieve high accuracy in electronic structure modeling, one should explicitly treat several (more than one) subvalence atomic shells of a heavy atom with different principal quantum numbers (tiny-core RPPs). This seems to be hardly compatible with the widely used semilocal representation of the $\hat{U}$ operator, implying the use of the same effective potential $U_{lj}(r)$ for all shells with the same 
spatial and total one-electron angular momenta ($l$ and $j$ respectively):
\begin{equation}
\hat{U} = \sum_{lj} U_{lj}(r) P_{lj},
\label{semilocal}
\end{equation}
where $P_{lj}$ projects onto the subspace of spinors with definite $l$ and $j$ values.
This restriction is quite acceptable for 
$s$- and $p$-elements, but is far from perfect for describing electronic structures of 
$d$- and especially 
$f$-element atoms and compounds where valence and subvalence shells are not well-separated spatially~\cite{Mosyagin:16,Mosyagin:17}. 
For excitation energies in actinide atoms and compounds, the errors arising from the use of  semilocal RPPs can reach 500 cm$^{-1}$ per each $f$-electron involved in the electronic transition. An efficient and general way to overcome the problem within the shape-consistent RPP framework is to use different partial potentials $U_{nlj}(r)$ for atomic shells with different principal quantum numbers $n$~\cite{Mosyagin:97,Titov:99}. Partial potentials 
${U}_{lj}$ in~(\ref{semilocal}) are 
replaced by the non-local operator
\begin{align}
\hat{U}_{lj} &= \sum_n \left[ U_{nlj}(r) P_{nlj}+P_{nlj} U_{nlj}(r) \right] \nonumber \\
      &-\frac{1}{2}\sum_{nn’}P_{nlj}\left[U_{nlj}(r)+U_{n’lj}(r)\right]P_{n’lj},
\label{eq:def-grpp}
\end{align}
where $P_{nlj}$
is a projector onto 
the subspace of 
subvalence atomic pseudospinors 
with quantum numbers $n$, $l$, and $j$.
In practice the operator~(\ref{eq:def-grpp}) is split into scalar-relativistic and spin-orbit parts. The presence of $P_{nlj}$ makes it a bit difficult to calculate integrals of the GRPP operators on the basis of atom-centered Gaussian functions. That is why most representative applications to date employing the 
full GRPP operator~(\ref{eq:def-grpp}) were restricted to diatomic molecules in quite modest basis sets and did not include any comprehensive calculations of actinide molecules (see~\cite{Mosyagin:01,Mosyagin:11,Mosyagin:13} and references therein). The general algorithm of evaluation of GRPP integrals in the molecular case was presented recently~\cite{Oleynichenko:LIBGRPP:23}; the integration of the non-local terms is even faster 
than the integration of the semi-local part. Pilot benchmark calculations of excitation energies of the ThO and UO$_2$ molecules have shown that the maximum deviation of GRPP from the four-component result does not exceed several dozens of wavenumbers. Given that pseudopotential also includes QED effects completely at no charge, one can argue that GRPPs can be regarded as one of the most precise relativistic 
Hamiltonians for molecular calculations (and one the least computationally demanding).

Generalized pseudopotentials are currently available for the whole Periodic table~\cite{GRPP_website}. GRPP integration engine~\cite{LIBGRPP:23} was interfaced to the DIRAC19 package~\cite{DIRAC_code:19,Saue:20} and is available under request.

\subsection{Intermediate Hamiltonian Fock-space relativistic coupled cluster theory with incomplete main model spaces}\label{sec:fsrcc}

One of the methods of 
solving
the many-electron problem most appropriate for theoretical supply of molecular spectroscopy is the relativistic version of the Fock-space multireference coupled cluster theory, FS RCC (for details see~\cite{Lindgren:87,Kaldor:91,Visscher:01,ShavittBartlett:09,Eliav:Review:22} and references therein). In the FS RCC framework exact electronic wavefunctions $\psi_i$ are expressed via the exponential wave operator $\Omega$ acting on model vectors $\tilde{\psi}_i$:
\begin{eqnarray}
\ket{\psi_i} = \Omega \ket{\tilde{\psi}_i},\quad\quad \Omega = \{\exp(T)\},
\label{eq:def-omega}
\end{eqnarray}
where $T$ stands for the cluster operator and curly braces mean that all contractions between $T$ operators are omitted. Cluster operator in the $n_hh\ n_pp$ Fock-space sector consists of contributions $T^{(n,m)}$ with $n$ valence hole and $m$ valence particle destruction operators:
\begin{equation}
T = \sum_{n\leq n_h} \sum_{m\leq n_p} T^{(n,m)},
\end{equation}
\begin{equation}
T^{(n,m)} = \sum_{LK} t_{LK}^{(n,m)} A_L^\dagger A_K,
\end{equation}
where the cluster amplitude $t_{LK}^{(n,m)}$ is associated with the excitation $A^\dagger_L A_K$, and $A_K$, $A^\dagger_L$ stand for chains of destruction and creation operators, respectively, defined with respect to the common Fermi vacuum determinant $\ket{\Phi_0}$. To find $t_{LK}^{(n,m)}$ one have to solve amplitude equations:
\begin{equation}
t^{(n,m)}_{LK} = D_{LK}^{-1} \left( V\Omega - \Omega (V\Omega)_{cl} \right)_{conn} \quad \forall A_K^\dagger\ket{\Phi_0},
\label{eq:ampl-equations}
\end{equation}
where $V = H - H_0$ stands for the perturbation operator, \textit{cl} marks the closed part of the operator and \textit{conn} denotes the connected part of the expression (in terms of Brandow diagrams). $D_{LK}$ is the conventional energy denominator associated with the excitation $K \rightarrow L$. In most practical applications $T$ includes only single and double excitations with respect to the model space determinants in the given sector (the FS RCCSD model). More sophisticated and computationally demanding, but much more accurate models include triple excitations partially (FS RCCSDT-n) or in a fully iterative way (FS RCCSDT)~\cite{Hughes:93,Musial:19,Oleynichenko:CCSDT:20}.

The conventional FS RCC method suffers from the intruder state problem~\cite{Evangelisti:87}, which manifests itself 
as
a presence of near-zero or positive denominators $D_{LK}$ leading to numerical instabilities arising during the iterative solution of Eqs.~(\ref{eq:ampl-equations}). To bypass this problem and ensure the smooth and stable behavior of calculated energies and properties in wide ranges of nuclear geometry parameters several approaches were proposed, e.~g. the intermediate Hamiltonian (IH)~\cite{Malrieu:85,Meissner:98,Landau:01,Eliav:XIH:05} and denominator shifting techniques~\cite{Zaitsevskii:RbCs:17,Zaitsevskii:Pade:18} (see also~\cite{Eliav:Review:22} for the recent review). In the present paper, we adopt the recent formulation of IH FS RCC based on the concept of an incomplete main model space (IMMS)~\cite{Zaitsevskii:QED:22}. Within this approach the whole model space $\mathcal{L}$ is split into the main subspace $\mathcal{L}_M$ nearly covering 
model-space parts of 
target electronic states and the intermediate subspace $\mathcal{L}_I$ serving as a buffer; in contrast with the previous formulation~\cite{Eliav:XIH:05},  $\mathcal{L}_M$ can be incomplete. The new formulation makes use of the correspondence of each excitation $A_L^\dagger A_K$ in Eq.~(\ref{eq:ampl-equations}) for any non-trivial sector to the sole determinant $A_K^\dagger \ket{\Phi_0}$ which does not vanish under the action of $A_L^\dagger A_K$. Cluster amplitudes associated with excitations corresponding to determinants belonging to $\mathcal{L}_M$ are calculated using the amplitude equations~(\ref{eq:ampl-equations}) with unmodified energy denominators,
whereas for those corresponding to
intermediate-space determinants, 
the
denominators are shifted by some quantities $S_{LK}$ in order to suppress intruder states:
\begin{align}
t^{(n,m)}_{LK} &= \left( D_{LK} + S_{LK} \right)^{-1} \left( V\Omega - \Omega (V\Omega)_{cl} \right)_{conn} \nonumber \\
& A_K^\dagger\ket{\Phi_0} \in \mathcal{L}_I.
\label{eq:ampl-equations-shifted}
\end{align}
In most practical cases the main model space $\mathcal{L}_M$ is readily defined based on some preliminary information on the electronic structure of target states $\psi_i$.
The shift parameter $S_{LM}$ can be set 
based on clear physical considerations~\cite{Zaitsevskii:QED:22}. 
Normally no additional parameters except for the definition of the main model space have to be specified, and the method works in a ``black-box'' manner. It was shown that in the case of enough large intermediate spaces calculated energies are very stable with respect to shift parameters.
The IMMS version of IH FS RCC is implemented in the EXP-T program package~\cite{Oleynichenko:EXPT:20,EXPT:23}.

\subsection{Direct 
evaluation
of transition property matrix elements 
}\label{sec:tranmom}

Despite multiple definitions of an effective property operator are possible~\cite{Hurtubise:93}, here we adopt that which seems to be the most natural for the Bloch formalism of effective operators underlying the FS RCC method (see~\cite{ShavittBartlett:09,Eliav:Review:22} and references therein). Within this formalism in addition to the wave operator $\Omega$ defined by the relation~(\ref{eq:def-omega}), one can also define the inverse mapping $\tilde{\Omega}$ 
\begin{equation}
    \bra{\psi_i} = \bra{\tilde{\psi}_i^{\perp\perp}} \tilde{\Omega},
\end{equation}
where $\tilde{\psi}_i^{\perp\perp}$ stands for the left model vector. Provided that model vectors are biorthonormalized, $\braket{\tilde{\psi}^{\perp\perp}_i|\tilde{\psi}_j} = \delta_{ij}$, property matrix element $O_{if}$ for the pair of electronic states $\psi_i$ and $\psi_f$ can be calculated via the relation~\cite{Hurtubise:93}:
\begin{equation}
    O_{if} = \braket{\tilde{\psi}_i^{\perp\perp}|\tilde{\Omega}O\Omega|\tilde{\psi}_f}\cdot N_i N_f^{-1},
\label{eq:def-eff-oper}
\end{equation}
where the normalization factors are defined as
\begin{equation}
    N_i = \braket{\psi_i|\psi_i}^{1/2} = \braket{\tilde{\psi}_i|\Omega^\dagger\Omega|\tilde{\psi}_i}^{1/2}
\end{equation}
(the same for $N_f$). This definition of the effective property operator leads to the non-Hermitian property matrix, $O_{if} \neq O_{fi}^*$ (in contrast to the alternative definition $ O_{if} = \braket{\tilde{\psi}_i|\Omega^\dagger O\Omega|\tilde{\psi}_f}\cdot N_i^{-1} N_f^{-1}$ which is inherently Hermitian~\cite{Hurtubise:93}). However, it presents no serious difficulty since any hermitization procedure can be applied. In particular, since $E1$ transition probabilities depend on squared matrix elements $|O_{if}|^2$, it can be beneficial to calculate directly this quantity since the normalization factors in~(\ref{eq:def-eff-oper}) will cancel each other:
\begin{equation}
    |O_{if}|^2 = O_{if} O_{fi} =
    \braket{\tilde{\psi}_i^{\perp\perp}|\tilde{\Omega}O\Omega|\tilde{\psi}_f}
    \braket{\tilde{\psi}_f^{\perp\perp}|\tilde{\Omega}O\Omega|\tilde{\psi}_i}.
\label{eq:hermitization}
\end{equation}
The latter formula closely resembles that widely used in the EOM-CC theory which also gives non-Hermitian property matrices~\cite{Jagau:16}.

Substituting the
well-known relation:
\begin{equation}
    \tilde{\Omega} = (\Omega^\dagger \Omega)^{-1} \Omega^\dagger,
\end{equation}
where the inversion of the $\Omega^\dagger\Omega$ operator is performed within the model space, 
and ``cutting'' the result by inserting the model-space projector $P = \sum\limits_m\ket{\tilde{\psi}_m}\bra{\tilde{\psi}_m^{\perp\perp}}$ we arrive at the working expression for property operator matrix elements:
\begin{align}
    O_{if} &= N_i N_f^{-1} \braket{\tilde{\psi}_i^{\perp\perp}|
    (\Omega^\dagger \Omega)^{-1} \Omega^\dagger O\Omega
    |\tilde{\psi}_f} = \label{eq:eff-oper-working} \\
    &= N_i N_f^{-1} \sum\limits_m\braket{\tilde{\psi}_i^{\perp\perp}|
    (\Omega^\dagger \Omega)^{-1}|\tilde{\psi}_m}
    \braket{\tilde{\psi}_m^{\perp\perp}|\Omega^\dagger O\Omega
    |\tilde{\psi}_f} \nonumber
\end{align}
To obtain matrix elements $\braket{\tilde{\psi}_i^{\perp\perp}| (\Omega^\dagger \Omega)^{-1}|\tilde{\psi}_m}$ one should simply calculate metric matrix $\braket{\tilde{\psi}_i^{\perp\perp}| \Omega^\dagger \Omega |\tilde{\psi}_m}$. Further inversion of this square matrix is always possible since it is never singular. Note that $\tilde{\psi}_i^{\perp\perp}$ and $\tilde{\psi}_m$ vectors always belong to the same irreducible representation and hence metric matrix is block diagonal.

The last but the most difficult point is to evaluate operators $\Omega^\dagger\Omega$ and $\Omega^\dagger O \Omega$ arising in~(\ref{eq:eff-oper-working}). The particular expressions for them depend on the coupled cluster Ansatz used. In the special case of FS CC, the normal-ordered exponential parametrization~(\ref{eq:def-omega}) of the wave operator is used, leading to the non-terminating series
\begin{equation}
    \Omega^\dagger \Omega = \{ e^{T^\dagger} \} \{ e^{T} \}
    \quad \text{and} \quad
    \Omega^\dagger O \Omega = \{ e^{T^\dagger} \} O \{ e^{T} \}
\label{eq:inf-sum}
\end{equation}
which have to be somehow artificially truncated. Here we propose to expand 
the right hand sides in~(\ref{eq:inf-sum}) in powers of $T$ and retain only the terms which are at most quadratic in $T$. Thus the expression for the ``metric'' term would be
\begin{equation}
    \Omega^\dagger\Omega \approx 1 + (T)_{cl} + (T^\dagger)_{cl} + (T^\dagger T)_{cl},
\label{eq:trunc-overlap}
\end{equation}
where the $_{cl}$ index stands for ``closed'' part of the operator. Linear terms have to be accounted for in the $1h1p$ sector, where a closed part of a cluster operator is non-zero, but are absent in the special case of purely particle sectors like $0h2p$ discussed in the present paper. The analogous expression for the property part is:
\begin{equation}
    \Omega^\dagger O \Omega \approx
    O + T^\dagger O + O T + \frac{(T^\dagger)^2}{2} O + T^\dagger O T + O \frac{T^2}{2}.
\label{eq:trunc-prop}
\end{equation}

It is natural to use the same level of truncation for both terms in~(\ref{eq:inf-sum}). In particular, it seems to be quite consistent to omit the normalization factors (quadratic in $T$) completely if the linear approximation is used for the property term, $\Omega^\dagger O \Omega \approx O + T^\dagger O + O T$. In principle, the quadratic truncation may be insufficient if some cluster amplitudes are large enough, and one can expect that even fourth-order contributions would be non-negligible for high precision in some cases (like it was shown for the expectation value calculations in the $0h0p$ sector~\cite{Noga:88}). However, even the cubic approximation leads to the explosive growth in the number of Brandow diagrams representing the terms in~(\ref{eq:inf-sum}), making the problem intractable (especially at the FS CCSDT level). Note that~(\ref{eq:inf-sum}) includes both connected and disconnected terms, which also have to be calculated and accounted for. However, it could be shown (see Appendix A) that for the special case of the quadratic truncation, disconnected terms in $\Omega^\dagger\Omega$ and $\Omega^\dagger O \Omega$ \textit{approximately} cancel each other resulting in the fully connected expression.

Note that our approach treats cluster operators from all Fock space sectors on equal footing. The alternative approach based on the separation of the vacuum sector amplitudes is more popular (see, for example,~\cite{Gopakumar:02}). In this case, the transformed property operator $(e^{T^{0h0p}})^\dagger O e^{T^{0h0p}}$ is built at the first step and then is contracted with amplitudes from non-trivial sectors. However, three-body terms of $(e^{T^{0h0p}})^\dagger O e^{T^{0h0p}}$ (and represented by six-index arrays) inevitably arise. To our best knowledge, they are quietly thrown away without any physical reason in actual program implementations, and the internal consistency of the overall scheme suffers.


\section{Computational details}\label{sec:comp-details}

The GRPP-based electronic structure model and FS RCC computational procedure employed in the present work essentially coincides with that described in Ref.~\cite{Oleynichenko:LIBGRPP:23}. The GRPP incorporating Breit and QED effects replaced 28 inner-core electrons of Th, whereas for the oxygen
atom we adopted the empty-core model~\cite{Mosyagin:21} (all electrons are retained, GRPP only simulates relativistic effects). The Fock space scheme ThO$^{2+}(0h0p)$ $\to$ ThO$^{+}(0h1p)$ $\to$ ThO$^{0}(0h2p)$ was assumed. All explicitly treated electrons except for those of innermost shells ($4spdf$ Th and $1s$ O) were correlated.
In studies of
the dependencies of calculated quantities on the internuclear separation $r$  we used the primitive $(19s\,17p\,15d\,15f)$ Th Gaussian set augmented with a contracted component (atomic natural orbitals $(7g\,6h\,5i)/[5g\,4h\,3i]$) ~\cite{Oleynichenko:LIBGRPP:23} and the standard aug-cc-pVQZ-DK set for O~\cite{Dunning:89,Kendall:92,Jong:01}. The FS RCC cluster operator expansion was restricted to single and double excitations (FS RCCSD); the model space at the FS RCC stage was somewhat larger than in Ref.~\cite{Oleynichenko:LIBGRPP:23} (35 Kramers pairs of active molecular spinors). The incomplete main model space for the target sector comprised all distributions of active electrons among 6 lowest-energy pairs of active spinors plus all single excitations out of this subset to one of the seven subsequent spinors. For all 19 states with equilibrium term energies ($T_e$) below 20 000 cm$^{-1}$ and the whole $r$ range considered (1.628 -- 2.158~\AA),  the
fractions of main model space determinants in the model vectors
were always larger than 95\%. 

Following the scheme described in Refs.~\cite{Isaev:21,Zaitsevskii:RaF:22} (see also Ref.~\cite{Pazyuk:15}), we used the single-reference relativistic coupled cluster method with the perturbative account of the contribution from triple excitations (RCCSD(T)) for ground-state energy calculations. Excited state energies as functions of the internuclear separation $r$ (and of the external field strength if needed) were obtained by combining the FS RCCSD electronic excitation energies and RCCSD(T) ground state energies. The resulting potentials which will be labeled as FS RCCSD / RCCSD(T) were used to evaluate numerically
energies and wavefunctions of the three lowest vibrational states of each term and derive the corresponding vibrational constants $\omega_e$.


In order to reduce the effect of basis set restriction on calculated term values we recomputed vertical excitation energies at $r= 1.864$~\AA{} (this value is quite close to equilibrium separations of all states under study) with an extended basis set obtained from the original one by adding additional single sets of functions $s$ through $i$ on Th and replacing the oxygen basis by aug-cc-pV5Z-DK without $h$ functions. The corrections thus obtained, $T_{bas}$, were added to $T_e$ values.

Molecule frame dipole moment values as functions of $r$ were calculated with the help of the conventional finite-field technique.
The radiative decay rates of excited rovibrational states were evaluated according to the Tellinghuisen’s sum rule~\cite{Tellinghuisen:84}. 
The required expectation values were calculated with vibrational functions, corresponding to FS RCC/RCCSD(T) potential and, whenever possible, to empirical (Rydberg--Klein--Rees, RKR) potentials.

The FS RCC calculations were performed with the EXP-T code~\cite{Oleynichenko:EXPT:20,EXPT:23}. The DIRAC19 program suite~\cite{DIRAC_code:19,Saue:20} interfaced to LIBGRPP library~\cite{Oleynichenko:LIBGRPP:23} was used to solve relativistic Hartree-Fock equations and obtain transformed molecular integrals.
The DIRAC19 code was also employed for 
single-reference
RCCSD(T) calculations. Vibrational energies and wavefunctions were evaluated with the help of the VIBROT program~\cite{Sundholm}.  RKR potentials were derived from available spectroscopic constant with the help of the code by A. Stolyarov.


\section{Results and discussion}\label{sec:results}

\begin{figure}[htp]
\centering
\includegraphics[width=0.9\columnwidth]{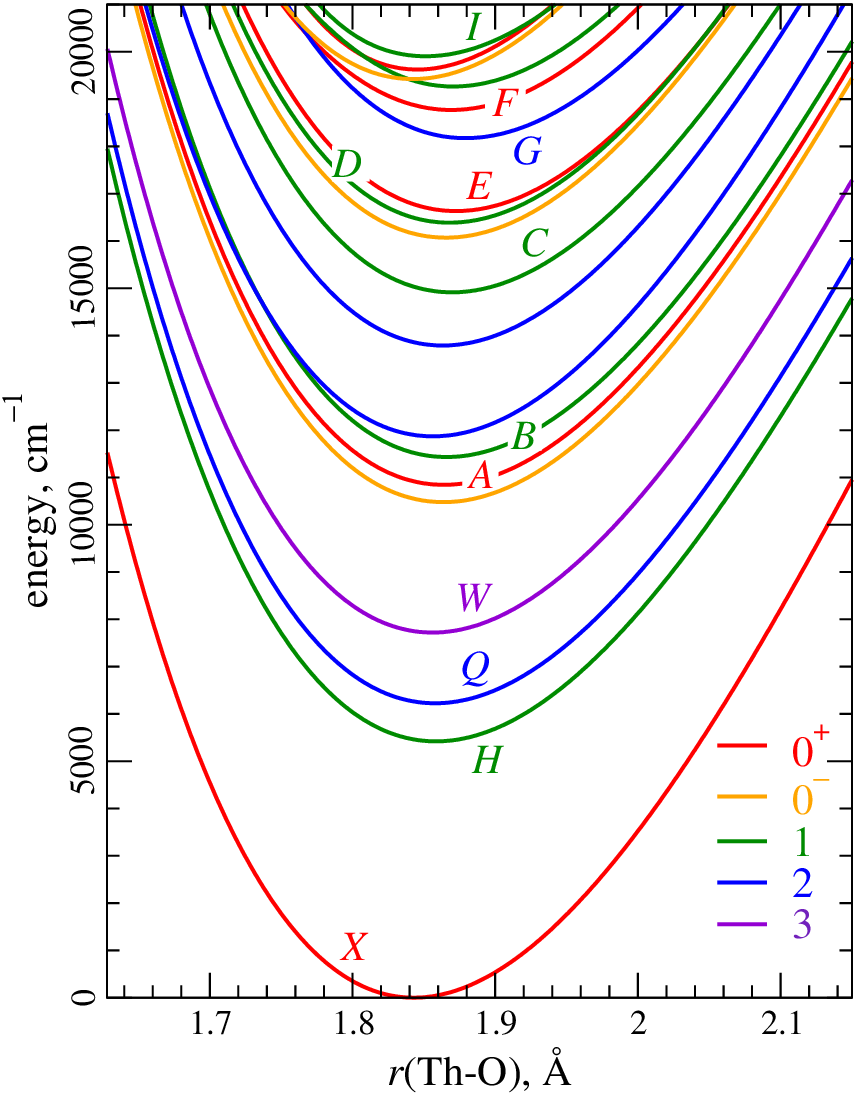}
\caption{FS RCCSD/RCCSD(T) adiabatic potential energy functions of low-lying electronic states of ThO. }    
\label{figun}
\end{figure}

\begin{figure}[htp]
\centering
\includegraphics[width=0.9\columnwidth]{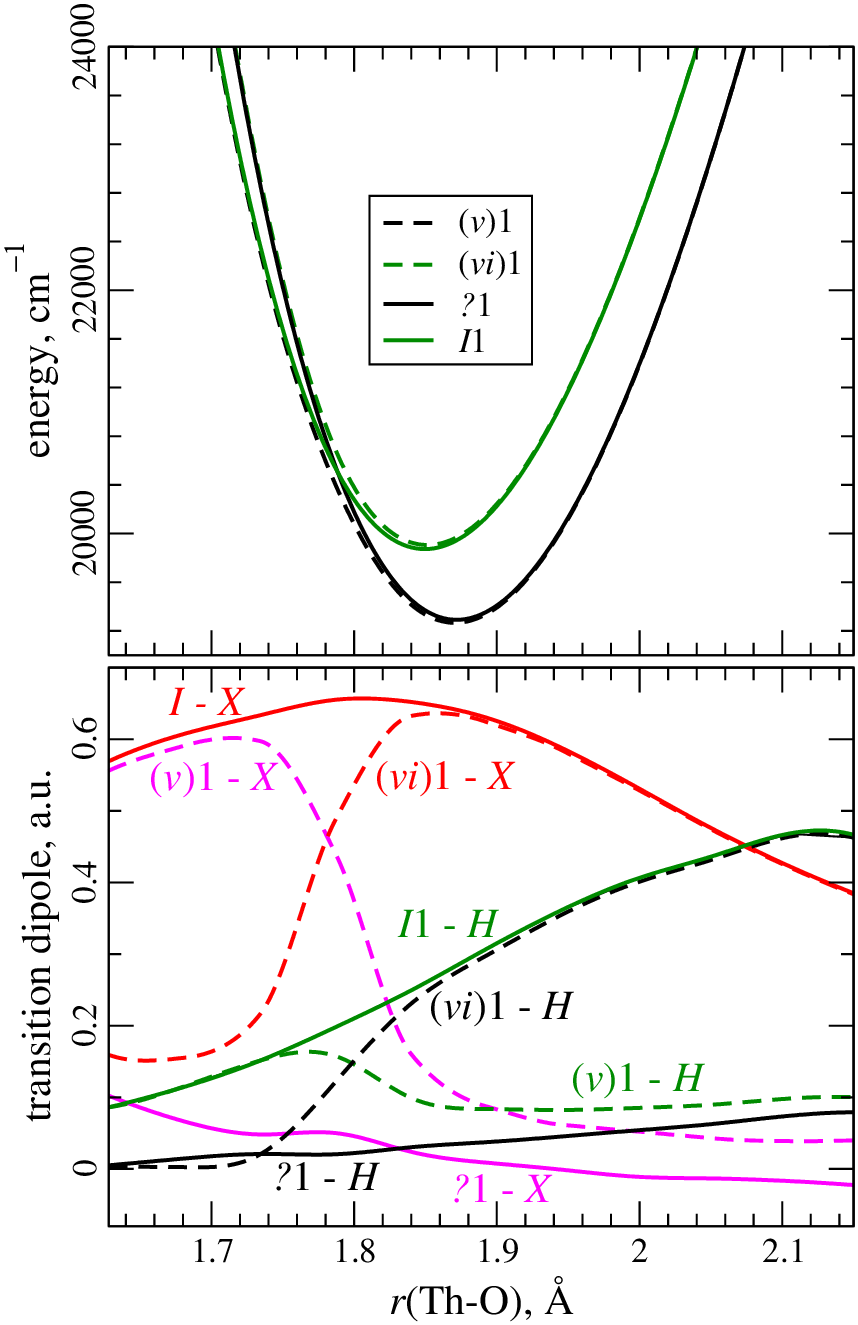}
\caption{Adiabatic (dashed lines) and quasidiabatic  (solid lines) potential energy functions for $(v,vi)1$ states of ThO and dipole moment functions for the $(v,vi)1\to X\,0^+$ and $(v,vi)1\to H\,1$ transitions. }
\label{figdeux}
\end{figure}

\paragraph*{Potential energy functions and spectroscopic constants.
}
%
The calculated potential energy functions are plotted in Fig.~{\ref{figun}}. In most cases, the assignment of the resulting adiabatic states to their spectroscopic analogs was straightforward. An important exception was the case of the sixth state with $\Omega=1$. The avoided crossing of the $(vi)\,1$ and $(v)\,1$ potential energy curves rather close to the minimum point of the former one (Figs.~\ref{figun} and \ref{figdeux}) puts into question the sense of single-electronic-state approximation for the corresponding vibronic states. Due to different magnitudes of errors for the states with different physical natures, the task of accurate non-adiabatic treatment of these vibronic states basing exclusively on the present results of electronic structure modeling seems unrealistic. 
In such a situation, it is reasonable to try to associate the spectroscopic electronic states with (quasi)diabatic states.  We used the naive two-state quasidiabatization scheme described in Ref.~\cite{Lefebvre:04}, approximating the  $r$-dependence of the rotation angle $\theta$ defining the 2$\times{}$2 transformation from adiabatic to quasidiabatic electronic states by a parametric function
\begin{equation}
\theta(r)=\frac{\pi}{2}{\rm arccot}\left(\frac{-a_x(r-r_x)}{\Delta_x}\right)
\label{lefebvre}
\end{equation}
where $r_x$ is the crossing point of quasidiabatic potentials whereas $\Delta_x$ and $a_x$ denote respectively the difference between the adabatic potentials and the slope of the difference between quasidiabatic potentials at $r=r_x$.
One of the quasidiabatic states can be identified with the spectroscopic $I\,1$ one; we shall denote the second state by $?\,1$ (see Fig.~\ref{figdeux}).

The molecular constants derived from the FS RCCSD/RCCSD(T) potential energy functions along with their experimental counterparts and the corresponding results of previous theoretical studies are collected in Tables~\ref{molconst},~\ref{molconsta}. The amount of theoretical data on excited electronic states of ThO is huge; here we restrict our attention to apparently most accurate data obtained in all-electron intermediate-Hamiltonian FS RCCSD calculations~\cite{Tecmer:18} performed with the Dirac-Coulomb
Hamiltonian and complete main model space formalism. For the states which were not accessible in the mentioned study~\cite{Tecmer:18}, Table~\ref{molconst} provides molecular constants obtained
in the framework of
the multireference second-order perturbation theory and Douglas-Kroll third-order relativistic two-component Hamiltonian~\cite{Paulovic:03}. For the two lowest states, we also cite the results of high-level single-reference relativistic coupled-cluster calculations \cite{Skripnikov:16,Smirnov:20}.

\begin{table*}[htp]
\caption{
Term energies $T_e$, equilibrium internuclear separations $r_e$, vibrational constants $\omega_e$ and molecule frame dipole moment values $D$ for low-lying $|\Omega|=0$ electronic states of ThO. PW: present FS RCCSD / RCCSD(T) calculations; $T_e$ include $\Delta T_{bas}$ corrections. Unless otherwise indicated, experimental molecular constants are extracted from Ref.~\cite{Edvinsson:85jms} and dipole moments are calculated for the lowest vibrational state ($v=0$). Two $D$ estimates separated by a slash correspond to vibrational solutions obtained with the RKR and present FS RCCSD/RCCSD(T) potentials respectively.       
}
\begin{center}
\renewcommand{\arraystretch}{1.2}
\begin{tabular*}{\textwidth}{r@{\extracolsep{\fill}}lcccc} 
\hline                   &        &$T_e,$ cm$^{-1}$&  $r_e$, \AA              &  $\omega_e$, cm$^{-1}$  &  $-D$, Debyes \\
\hline $X\,(i)0^+$        &PW      &  0             &        1.843             & 898                     &   2.753/2.776     \\
                   &Exptl.  &  0             &        1.840             & 896                     &  2.782$\pm{}$0.012$^{\,a}$\\
                   &Theor.   &  0             &1.837$^{b}\!\!$, 1.841$^{c,d}$& 922$^{\,b,c}$, 897$^{d}$ & 2.93$^{\,e}$\\[1ex]
$A\,(ii)0^+$       &PW      &  10 847      &  1.864             & 853                 & 1.849/1.811     \\
                   &Exptl.  &  10 601        &        1.867             & 846 \\
                   &Theor.   &  11 699$^{\,b}\!\!$, 11 292$^{\,c}$ & 1.852$^{\,b}\!\!$, 1.867$^{\,c}$& 910$^{\,b}\!\!$, 882$^{\,c}$\\[1ex]  
$E\,(iii)0^+$      &PW      &  16 567&        1.873             & 810                 & 3.401/3.422 \\
                   &        &                    &                          &                     & $(v\!=\!1)$ 3.448/3.468 \\
                   &Exptl.  &  16 320            &        1.867             & 829  & $(v\!=\!1)$ 3.534$\pm{}$0.010$^{\,a}$ \\
                   &Theor.   &  14 370$^{\,b}\!\!$, 17 280$^{\,c)}$ & 1.868$^{\,b}\!\!$, 1.859$^{\,c}$& 855$^{\,b}\!\!$, 875$^{\,c}$\\[1ex]  
$F\,(iv)0^+$       &PW      &  18 685&        1.869             & 808  &     --- /4.621 \\
                   &Exptl.  &  18 338            &        1.870             & 758 \\[1ex]
$(v)0^+$           &PW      &  19 623&  1.845                   & 911  &    --- /4.754 \\[2ex]
$(i)0^-$           &PW      & 10 486 &  1.865                   & 853  & --- /1.585\\
                   & Theor.  & 10 701$^{\,b}\!\!$, 10 911$^{\,c}$&1.861$^{\,b}\!\!$, 1.857$^{\,c}$&857$^{\,b}\!\!$, 882$^{\,c}$ \\[1ex]
$(ii)0^-$          &PW      & 16 026 &  1.865                   & 846  & --- /1.339 \\
                   &Theor.  &  16 982$^{\,b}\!\!$, 18 016$^{\,c}$& 1.888$^{^{\,b}\!\!}$, 1.868$^{\,c}$ & 822$^{\,b}\!\!$, 855$^{\,c}$  \\[1ex]
$(iii)0^-$         &PW      & 19 438 &  1.843                   & 871  & --- /4.699 \\
\hline
\end{tabular*}   
\end{center}
$^a$) Ref. \cite{Wang:11}; $^b$ and $^{c}$) intermediate-Hamiltonian all-electron FS RCC calculations \cite{Tecmer:18} with ThO and ThO$^{2+}$ Fermi vacuum states respectively; $^{d}$) composite single-reference coupled-cluster scheme accounting for triples and perturbative quadruples \cite{Smirnov:20}; $^e$) all-electron single-reference RCCSD(T) \cite{Buchachenko:10}
\label{molconst}
\end{table*}

\begin{table*}[htp]
\caption{
Term energies, equilibrium internuclear separations, vibrational constants, and molecule frame dipole moment values for low-lying electronic states of ThO with $|\Omega|\ge 1$. See Table \ref{molconst} for notation and explanations.       
}
\begin{center}
\renewcommand{\arraystretch}{1.2}
\begin{tabular*}{\textwidth}{r@{\extracolsep{\fill}}lcccc}
\hline                   &        &$T_e,$ cm$^{-1}$&  $r_e$, \AA              &  $\omega_e$, cm$^{-1}$  &  $-D$, Debyes \\
\hline $H\,(i)1$          &PW      &  5 391 &        1.859             & 863   & 4.126/4.132                    \\
                   &Exptl.  &  5 317             &        1.858             & 857   & 4.24$\pm $0.10$^{\,f}\!\!$, 4.25$\pm$0.02$^{\,g}$\\  
                   & Theor.  &  5 168$^{\,b}\!\!$, 6 017$^{\,c}$,    
                                                 & 1.854$^{\,b}\!\!$, 1.855$^{\,c}$& 885$^{\,b,c}$ & 4.24$^{\,h}$\\
                   &        &    5 327$^{g}$ \\[1ex]
$B\,(ii)1$         &PW      &  11 429&        1.866             & 850 & 1.769/1.793 \\
                   &Exptl.  &  11 129            &        1.864             & 843 \\
                   &Theor.   &  12 056$^{\,c}$     & 1.859$^{\,c}$           & 879$^{\,c}$\\[1ex]
$C\,(iii)1$        &PW      &  14 889&        1.870             & 843 &  2.526/2.518\\ 
                   &Exptl.  &  14 490            &        1.870             & 825 &  2.60$\pm$0.04$^{\,i}$\\
                   &Theor.   &  14 451$^{\,b}\!\!$, 16 188$^{\,c}$ &  1.866$^{\,b}\!\!$, 1.864$^{\,c}$ & 859$^{\,b}\!\!$, 869$^{\,c}$      \\[1ex]
$D\,(iv)1$         &PW      &  16 345&        1.868             & 845 & 1.813/1.832\\
                   &Exptl.  &  15 946            &        1.866             & 839 \\
                   &Theor.   &  17 644$^{\,c}$   & 1.862$^{\,c}$ & 874$^{\,c}$      \\[1ex]
$?\,(v,\,vi)1$     &PW      &  19 187&        1.873             & 832 & --- /5.683$^{\,j}$\\             
$I\,(v,\,vi)1$     &PW      &  19 854&        1.850             & 824 &  4.246/4.248$^{\,j}$ \\            
                   &Exptl.  &  19 539            &        1.849             & 801 \\[2ex]
$Q\,(i)2$          &PW      &  6 192  &        1.858             & 863 &   4.036/ 4.051     \\      
                   &Exptl.  &  6 128             &        1.856             & 858 & 4.07$\pm$0.06$^{\,i}$  \\
                   &Theor.   &  6 086$^{\,b}\!\!$, 6 866$^{\,c}$ & 1.853$^{\,b}\!\!$, 1.854$^{\,c}$ & 886$^{\,b,c}$&  \\[1ex]
$(ii)2$            &PW      &  11 818&        1.856             & 859  & --- /2.855                              \\            
                    &Theor.  &  12 803$^{\,b}\!\!$, 12 732 $^{\,c}$& 1.849$^{\,b}\!\!$, 1.852$^{\,c}$ & 885$^{\,b}\!\!$, 886$^{\,c}$      \\
$(iii)2$           &PW      &  13 765&        1.863             & 855  & --- /2.043                              \\            
                    &Theor. &  14 997$^{\,b}\!\!$, 14 553 $^{\,c}$& 1.859$^{\,b}\!\!$, 1.857$^{\,c}$ & 872$^{\,b}\!\!$, 883$^{\,c}$       \\
$G\,(iv)2$         &PW      &  18 135&        1.879             & 823  &  3.254/3.228\\            
                   &Exptl.  &  18 010            &        1.882             & 809  \\
                   &Theor.  & 17 339$^{\,k}$      &        1.920$^{\,k}$      & 759$^{\,k}$  \\[2ex]
$W\,(i)3$          &PW      &   7 660&        1.856             & 865  & --- /4.095\\
                   &Exptl.  &   8 600$^{\,l}$ (?)&                          &      \\           
                   &Theor.   &7 694$^{\,b}\!\!$, 8 438 $^{\,c}$ & 1.852$^{^{\,b,c}}$ & 887$^{\,b}\!\!$, 889$^{\,c}$ \\
\hline
\end{tabular*}   
\end{center}
$^{f}$) Ref.~\cite{Vutha:11}; $^{g}$) Ref.~\cite{Kokkin:15}; $^{h}$)  four-component single-reference RCC \cite{Skripnikov:16}; $^{i}$)\cite{Wu:20}; $^{\,j}$) estimated using the interpolation of adiabatic dipole moment values at large distances from the avoided crossing point;  $^{k}$) all-electron second-order multireference perturbation theory calculations \cite{Paulovic:03}; $^l$) estimate taken from \cite{Kuchle:94}
\label{molconsta}
\end{table*}

The deviations of the present $T_e$ estimates from their well-established spectroscopic counterparts never exceed 400 cm$^{-1}$ (rms deviation 280 cm$^{-1}$); it is to be
emphasized
that the error is always smaller than a half of vibrational quantum for the corresponding state. The corrections $\Delta T_{bas}$
were normally moderate (several dozens of wavenumbers) and improved the agreement between the theoretical and experimental values; the largest correction (104 cm$^{-1}$) concerns the $(v)1$ state. Our results do not confirm the empirical estimate $T_e=8 600$ cm$^{-1}$ for the $W\,(i)3$ state appeared in Refs.~\cite{Paulovic:03,Tecmer:18}. The large error in our value, 7 660 cm$^{-1}$, seems hardly probable because of nearly perfect reproduction of spectroscopic constants for other components of the lowest $^3\Delta$ manifold, $H1$ and $Q2$. The present calculations reproduce correctly rather subtle variations of equilibrium internuclear separations from state to state; a relatively large deviation from the experimental $r_e$ (+0.006~\AA{}) was found only for the  $E\,0^+$ state. The discrepancies between the FS RCCSD/RCCSD(T) and experimental vibrational constants are normally well within two dozens of wavenumbers; the exceptional case of the $F\,(iv)0^+$ state should be noticed (the calculated $\omega_e$, 808 cm$^{-1}$, differs significantly from the spectroscopic value, 758 cm$^{-1}$). Radical improvements in accuracy (normally 3 to 4 times for $T_e$) over the previous FS RCC calculations with similar choice of Fock space scheme and total model space can be explained mainly by employing a more adequate approximation for the many-electron Hamiltonian incorporating the bulk of Breit effects (up to 600 cm$^{-1}$ for excitation energies, according to Ref.~\cite{Oleynichenko:LIBGRPP:23}). The use of incomplete main model spaces is essential for describing the states in the upper part of the studied energy range.

It is believed instructive to analyze the interrelationship between
relativistic molecular electronic states and their scalar ($\Lambda$--$S$) counterparts. To this end, we repeated FS RCCSD calculations for some averaged equilibrium value of $r$ (1.864 \AA) with
spin-orbit parts of
pseudopotentials
nearly
switched off, and projected the model-space parts of the obtained scalar relativistic states onto those of fully relativistic states (cf.~\cite{Zaitsevskii:RbCs:17}). The results are summarized in Table~\ref{compo}. For 14 lowest states, these results generally align with those from Ref.~\cite{Paulovic:03}. Taking into account that the angle $\theta{}$ in Eq.~(\ref{lefebvre}) at the assumed $r$ value is about $78^\circ$ and corresponds to ca. 96:4 state mixing, one can conclude that the $I\,1$ state is mainly a $^3\Pi$--$^1\Pi$ mixture with a certain predominance of the $^3\Pi$ component whereas the other quasidiabatic state, $?1$, is strongly dominated by the contribution from $^3\Sigma^-$. 

\begin{table}
\caption{Composition of low-lying relativistic electronic states of ThO in terms of scalar relativistic ones ($r=1.864$~\AA). Contributions below 2\% are not shown.}
{
\renewcommand{\arraystretch}{1.2}
\begin{tabular*}{\columnwidth}{l@{\extracolsep{\fill}}ll}
\hline State & $\Omega$ & Composition \\
\hline $X\,(i)$ & $0^+$        & $99\% \,(i)^1\Sigma^+$\\
$A\,(ii)$ & $0^+$       & $90\%\,(i)^3\Pi$, $8\%\,(ii)^1\Sigma^+$\\
$E\,(iii)$ & $0^+$      & $72\%\,(ii)^1\Sigma^+$,  $18\%\,(i)^3\Sigma^-$, $6\%\,(i)^3\Pi$, $3\%\,(iii)^1\Sigma^+$ \\ 
$F\,(iv)$ & $0^+$       & $56\%\,(i)^3\Sigma^-$, $16\%\,(ii)^1\Sigma^+$, $14\%\,(ii)^3\Pi$,  $9\%\, (iii)^1\Sigma^+$, \\
          &             & $3\%\,(i)^3\Pi$  \\
$(v)$ & $0^+$           & $79\%\,(ii)^3\Pi$, $8\%\,(i)^3\Sigma^-$, $4\%\,(ii)^1\Sigma^+$, $3\%\,(iii)^3\Pi$, \\
      &                 & $2\%\,(iii)^1\Sigma^+$, $2\%\,(iv)^1\Sigma^+$ \\[2ex]
$(i)$ & $0^-$           & $83\%\,(i)^3\Pi$, $15\%\,(i)^3\Sigma^+$ \\ 
$(ii)$ & $0^-$          & $ 83\%\,(i)^3\Sigma^+$, $ 16\%\,(i)^3\Pi$ \\
$(iii)$ & $0^-$         &  $ 92\%\,(ii)^3\Pi$, $ 04\%\,(ii)^3\Sigma^+$ \\[2ex] 
$H\,(i)$ & $1$          & $ 99\%\,(i)^3\Delta$ \\
$B\,(ii)$ & $1$         & $ 69\%\,(i)^3\Pi$, $ 23\%\,(i)^3\Sigma^+$, $ 7\%\,(i)^1\Pi$   \\
$C\, (iii)$ & $1$        & $ 38\%\, (i)^1\Pi$, $ 32\%\, (i)^3\Sigma^+$, $ 27\%\, (i)^3\Pi$ \\ 
$D\, (iv)$ & $1$         &  $ 52\%\, (i)^1\Pi$, $ 44\%\, (i)^3\Sigma^+$, $ 2\%\, (i)^3\Pi$ \\
$(v)$ & $1$             & $ 89\%\, (i)^3\Sigma^-$, $ 7\%\, (ii)^3\Pi$, $ 2\%\, (ii)^1\Pi$ \\             
$(vi)$ & $1$            & $ 60\%\, (ii)^3\Pi$, $ 17\%\, (ii)^1\Pi$, $ 9\%\, (i)^3\Sigma^-$, $ 5\%\, (ii)^3\Delta$, \\
       &                & $ 3\%\, (ii)^3\Sigma+$, $ 2\%\, (i)^1\Pi$, $ 2\%\, (iii)^3\Pi$ \\[2ex]            
$Q\, (i)$ & $2$          &  $  93\%\, (i)^3\Delta$, $ 5\%\, (i)^1\Delta$ \\ 
$ (ii)$ & $2$           &  82$\%\, (i)^1\Delta$,  10$\%\, (i)^3\Pi$,  6$\%\, (i)^3\Delta$  \\            
$(iii)$ & $2$           & $ 88\%\,  (i)^3\Pi$, $ 11\%\, (i)^1\Delta$ \\         
$G\, (iv)$ & $2$         &  $ 89\%\, (i)^3\Phi$, $ 4\%\, (ii)^3\Delta$ \\[2ex]   
$W\, (iii)$ & $3$           &  100\% $(i)^3\Delta$ \\
\hline
\end{tabular*}
}
\label{compo}
\end{table}

\paragraph*{Molecule frame 
dipole moments.}

For most electronic states under study, molecule frame dipole moments rapidly and regularly increase within the whole range of internuclear separations considered (see Fig.~\ref{figtrois}; numerical data on all 19 states can be found in Supplementary materials). Irregular behavior of dipole moment functions for the adiabatic $(v)1$ and $(vi)1$ states is related to the avoided crossing discussed above. Expectation values of dipole moments for the lowest vibrational levels (and for the first excited level, the $E\,(iii)0^+$ state where the experimental counterpart is known) are listed in Tables~\ref{molconst},~\ref{molconsta}. Despite strong dependencies of dipole moments on $r$, which imply a high sensitivity of expectation values on the input data for the vibrational problem, the differences of these values computed with vibrational eigenfunctions of RKR and FS RCCSD/RCCSD(T) potentials are not significant. This fact confirms the reasonable accuracy of calculated potential curves indirectly. The resulting estimates are in a very good (within a few hundredths of a Debye) agreement with measured values for $X\,0^+$, $E\,0^+$, $C\,1$, and $Q\,1$ states; the discrepancy ca.~0.1 Debye is observed for the $H\,1$ state. It might be worth noting that within the present combined FS RCCSD/RCCSD(T) scheme the ground-state dipole moment values are simply RCCSD(T) ones; the difference from the CCSD(T) results from Ref.~\cite{Buchachenko:10} arises from the use of more accurate relativistic Hamiltonian and more flexible Th-centered part of the employed basis set.     

\begin{figure}[htp]
\centering
\includegraphics[width=0.9\columnwidth]{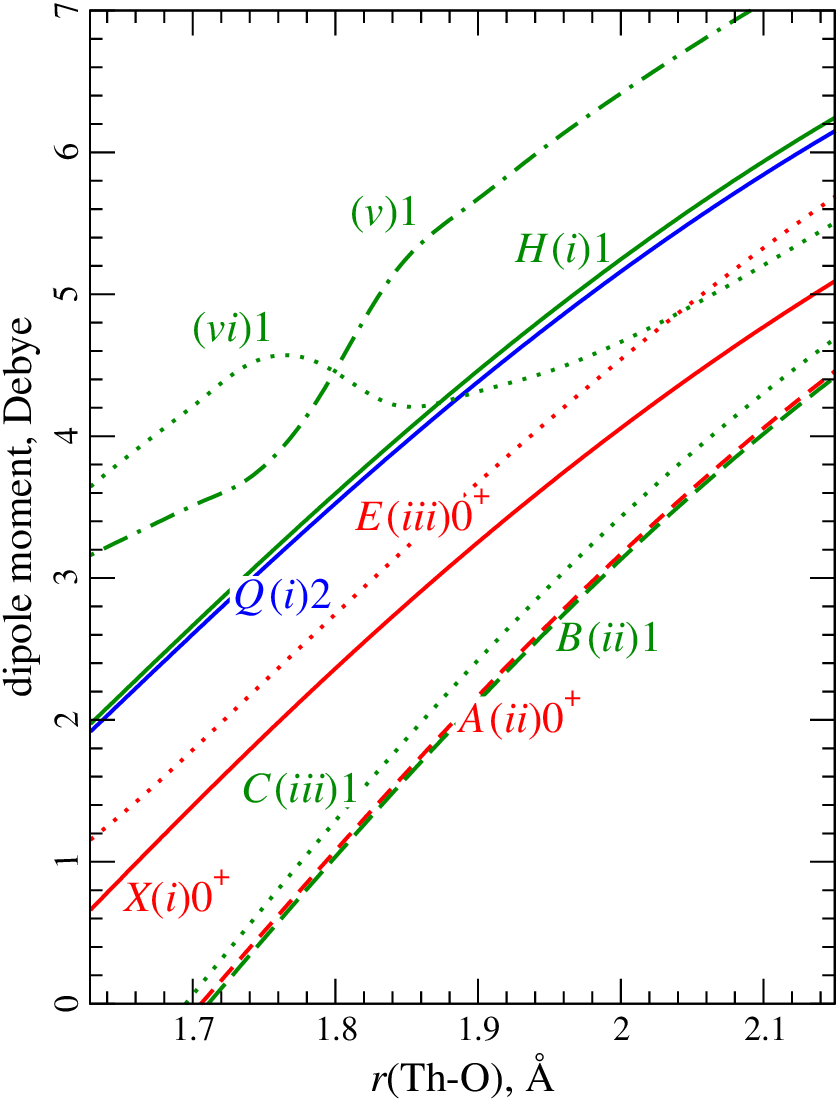}
\caption{FS RCCSD/RCCSD(T) molecule frame dipole moment functions for several low-lying adiabatic electronic states of ThO. Positive sign corresponds to Th$^{\delta+}$O$^{\delta-}$charge distributions.}
\label{figtrois}
\end{figure}

\paragraph*{Transition dipole moments and excited state lifetimes.} Due to significant gap between the three lowest states and other states with $|\Omega|\le 2$ as well as to moderate differences between the equilibrium bond lengths in different low-lying states, preliminary information on most probable radiative decay channels for the states within the energy interval 10 000 - 20 000 cm$^{-1}$ optically accessible from the ground one (i.e. with $|\Omega|\le 1$) can be readily obtained from vertical transition dipole moment values collected in Table~\ref{verttdm}; spontaneous decay of these states to other lower states is strongly suppressed by energy factor.

\begin{table}
\caption{FS RCC absolute values of transition dipoles (in atomic units) for vertical transitions in ThO from/to the ground state, $H\,1$ and $Q\,2$ states ($r= 1.864$~\AA).}
{
\renewcommand{\arraystretch}{1.2}
\begin{tabular*}{\columnwidth}{r@{\extracolsep{\fill}}ccc}
\hline                   &     $X\,(i)0^+\to$               &  $H\,(i)1\to$               &  $Q\,(i)2\to$   \\
\hline $X\,(i)0^+$        &       --                          &   0.031                  &  -- \\
$A\,(ii)0^+$       &     0.341                     &   0.528                  &  -- \\
$E\,(iii)0^+$      &     0.765                     &   0.024                  &  -- \\
$F\,(iv)0^+$       &     0.426                     &   0.019                  &  -- \\
$\quad (v)0^+$     &     0.102                     &   0.662                  &  -- \\[1ex]
$\quad (i)0^-$     &        --                         &   0.519                  &  -- \\
$\quad (ii)0^-$    &        --                         &   0.093                  &  -- \\
$\quad (iii)0^-$   &        --                         &   0.620                  &  --  \\[1ex]
$H\,(i)1$          &     0.031                     &       --                     & 0.073 \\
$B\,(ii)1$         &     0.431                     &   0.103                  & 0.510 \\
$C\,(iii)1$        &     0.624                     &   0.019                  & 0.395 \\
$D\,(iv)1$         &     0.477                     &   0.067                  & 0.117 \\
$?\,(v,\,vi)1$     &    0.0150            & 0.034          & 0.134 \\             
$I\,(v,\,vi)1$     &    0.644            & 0.275          & 0.176 \\[1ex]
$Q\,(i)2$          &             --                    &  0.073                   & -- \\
$ (ii)2$           &             --                    &  0.128                   & 0.145 \\            
$(iii)2$           &             --                    &  0.075                   & 0.054 \\
$G\,(iv)2$         &             --                    &  1.125                   & 0.277 \\[1ex]            
$ (i)\,3$          &             --                    &          --                  &  0.064 \\
\hline
\end{tabular*}
}
\label{verttdm}
\end{table}

\begin{figure}[htp]
\centering
\includegraphics[width=0.9\columnwidth]{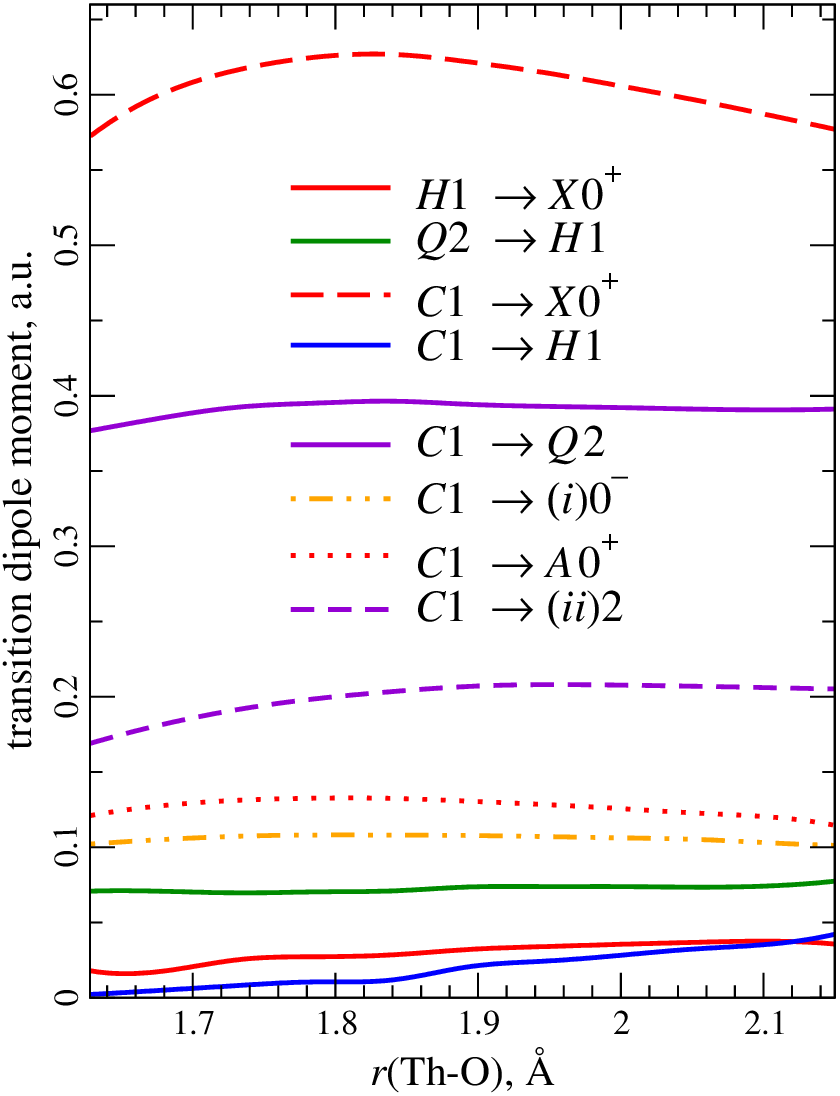}
\caption{FS RCCSD transition dipole moment functions for main radiative decay channels for the states $H1$, $Q1$, and $C1$.}
\label{figquatre}
\end{figure}

Since the main goal of the present study consists in assessment of the computational scheme 
outlined
above, we focus on describing the transitions which define the experimentally measured radiative lifetimes. Transition dipole moment functions for main decay channels for the states $H1$, $Q1$, $C1$, and $I1$ are presented in Figs. \ref{figquatre} and \ref{figdeux}. The corresponding estimates for partial and full radiative lifetimes of lowest vibrational levels ($v=0$) along with their measured counterparts are presented in Table~\ref{lifetimes}. A semiquantitative agreement between theoretical and available experimental lifetimes and branching ratios
is achieved in all cases. The computed radiative lifetime for the $H\,1$ state agrees well with experimental data, being somewhat shorter than its measured counterpart. Our results fully confirm the conclusion of Ref.~\cite{Wu:20} concerning very long radiative lifetime of the $Q2$ state. The $C1$ lifetime is only slightly shorter that the experimental value of Kokkin \emph{et al.}~\cite{Kokkin:14} and the branching ratio for the two main decay channels, $\to X0^+ $ and $\to Q2$, is correctly reproduced. Our estimate of
the $I\,1$ lifetime is ca.~20\% longer than the experimental one; the calculated $I1 \to X0^+ :H1:Q1$ branching ratio ($90:6:4$) seems
to be
fully compatible with the experimental ratio ($I1(v=0) \to X0^+(v=0,1):H1(v=0):Q1(v=0)$ = $92:5:3$)~\cite{Kokkin:14}. Among other things, this finding unambiguously confirms the correctness of identifying the $(v,vi)1$ quasidiabatic state with the spectroscopic $I$ one.  

\begin{table}[htp]
\caption{ 
Radiative lifetimes of several excited states of ThO derived from FS RCC transition dipole moments ($D^{\rm CC}$) and
FS RCCSD/RKR
or FS RCCSD/RCCSD(T) potential energy functions ($E^{\rm RKR}$ and $E^{\rm CC}$, respectively). Partial lifetimes for the channels with negligible contributions to the total decay rates are not shown.
}{
\renewcommand{\arraystretch}{1.2}
\begin{tabular*}{\columnwidth}{l@{\extracolsep{\fill}}ccc}
\hline                     &    $D^{\rm CC}$/$E^{\rm RKR}\,^{a)}$ & 
                                                         $D^{\rm CC}$/$E^{\rm CC}$ &      Exptl. \\
\hline $H\,(i)1$            &     3.83 ms      &   3.57  ms&   $\ge 1.8$   ms   \cite{Vutha:10} \\
& & &  4.2$\pm$0.5  ms  \cite{Ang:22}\\
$Q\,(i)2$            &     177 ms       &   182 ms  &  $ > 62$ ms \cite{Wu:20} \\
$C\,(iii)1$          &     400 ns          &  362  ns   &   468$\pm$30   ns   \cite{Kokkin:14}  \\
    &     &  & 468$\pm$30   ns   \cite{Kokkin:14}  \\
$\;\to X\,(i)0^+$    & 433 ns  &  393 ns \\
$\;\to Q\,(i)2$      &5.50 $\mu$s & 4.87 $\mu$s                                 & 5.4$\pm$1.3 $\mu$s$^{b)}$ \cite{Wu:20} \\
$\;\to (i)0^-$       &               & 175 $\mu$s   \\
$\;\to A\,(ii)0^+$   & 491 $\mu$s &  427 $\mu$s \\
$\;\to (ii)2$        &          &  428 $\mu$s \\
$I\,1\;\;\;$         &  141 ns          &  134 ns                 & 115$\pm$4   ns   \cite{Kokkin:14} \\
$\;\to X\,0^+$       &  161 ns          &  153 ns                 & 126 ns $^{b)}$ \cite{Kokkin:14}\\
$\;\to H\,1$         &  2.42 $\mu$s     &  2.29 $\mu$s            & 2.3 $\mu$s $^{b)}$ \cite{Kokkin:14}\\
$\;\to Q\,2$         &  3.40 $\mu$s     &  3.30 $\mu$s            & 3.8 $\mu$s $^{b)}$ \cite{Kokkin:14}\\
$\;\to B\,(ii)1$     & 16.1 $\mu$s          &  15.8 $\mu$s \\
$\;\to (ii)2$        &                  &  9.78 $\mu$s \\
\hline
\\
\end{tabular*}}
\\
$^a)$ Minor contributions from transitions to the states with missing spectroscopic data were taken from $D^{\rm CC}$/$E^{\rm CC}$; $^{b)}$ estimated from published transition moment values or/and branching ratios 
\label{lifetimes}
\end{table}


\section{Conclusion}\label{sec:conclusion}

Comprehensive theoretical study of excited states of the ThO molecule with term energies below 20 000 cm$^{-1}$ reported in the present paper is the first full-scale application of three tools for high-precision \textit{ab initio} modeling of excited electronic states of molecules developed by our group in the last few years: tiny-core generalized relativistic pseudopotentials accounting for Breit and QED effects, intermediate Hamiltonian FS RCC for incomplete main model spaces and the direct technique of transition property matrix elements evaluation, first introduced in this work. The results can be assessed as generally promising. The errors in predicting electronic term values are smaller than a half of vibrational quanta, a feature essential for correct interpretation of experimental spectra. Based on the obtained results one can argue that the present level of accuracy of transition moment calculations is sufficient for most applications.
The same applies for the molecule frame dipole moment. We expect that the possibility of purely theory-based identification of the strongest transitions will greatly simplify planning of future spectroscopic experiments with other short-lived radioactive molecules.

However, there are still unresolved challenges crucial for further progress of the theoretical molecular spectroscopy of actinide-containing molecules. First of all, the accuracy demonstrated in the present paper for term energies of ThO (rms error 280~cm$^{-1}$) seems to be nearly ultimate for the FS RCC approximation restricted to single and double excitations. Errors can become more substantial for molecules with more complicated electronic structure, and these errors cannot be reliably estimated and/or corrected without auxiliary calculations including corrections for higher excitations. The most important consequence of large error seems to be the impossibility of unambiguous numbering of experimentally measured vibrational progressions for electronic states with relatively soft vibrational modes. Thus the problem of construction of an effective computational scheme accounting for triple excitations for molecular problems with many hundreds of spinors involved in correlation treatment remains the most urgent challenge in the near future. The other purely ``technical'' problem is the lack of systematic sequences of basis sets adapted for use with GRPPs and allowing well-justified extrapolation to the basis set limit.

Last but not least, highly accurate treatment of electron correlation would bring to the fore the problem of solving a non-adiabatic vibrational problem. Due to high density of electronic states in actinide molecules, the development of such tools seem to be the next natural step towards quantitative modeling of rovibronic spectra for transitions involving such states. The systematic solution of this problem appeals for non-adiabatic coupling matrix element calculations. To date such calculations are not available in the relativistic coupled cluster framework, and the development of techniques aimed at these calculation is an intriguing and challenging task.

\section{Acknowledgements}

We are grateful to Leonid Skripnikov, Anatoly Titov, and Elena Pazyuk for useful discussions. Thanks are due to Andrey V. Stolyarov for supplying us with his code for constructing RKR potentials.
Electronic structure calculations have been carried out using computing resources of the federal collective usage center Complex for Simulation and Data Processing for Mega-science Facilities at National Research Centre ``Kurchatov Institute'', http://ckp.nrcki.ru/.

The development of new tools for transition property calculations at NRC ``Kurchatov Institute'' -- PNPI was supported by the Russian Science Foundation Grant No. 19-72-10019 (https://rscf.ru/project/22-72-41010/).

\bibliographystyle{apsrev}
\bibliography{ThO}

\end{document}